\documentclass[12pt]{iopart}

\usepackage{iopams}
\usepackage{multicol,mathrsfs,bm}
\usepackage{graphicx, graphics}
\usepackage{hyperref}
\usepackage{slashed}
\usepackage{fancyhdr}

\def\de{{\rm d}}

\newcommand{\gsim}{\gtrsim}
\renewcommand{\lsim}{\lesssim}
\def\gappeq{\mathrel{\rlap {\raise.5ex\hbox{$>$}} {\lower.5ex\hbox{$\sim$}}}}
\def\lappeq{\mathrel{\rlap{\raise.5ex\hbox{$<$}} {\lower.5ex\hbox{$\sim$}}}}
\def\beq{\begin{equation}} \def\eeq{\end{equation}}
\def\bea{\begin{eqnarray}} \def\eea{\end{eqnarray}}
\def\bq{\begin{quote}} \def\eq{\end{quote}}
\def\bc{\begin{center}} \def\ec{\end{center}}

\newcommand{\bkone}{{\rm K}_1}
\newcommand{\bktwo}{{\rm K}_2}

\begin{document}

\thispagestyle{empty}
  \begin{flushright}
 \textbf{JCAP} 08 (2007) 013\\
{\small MIT-CTP-3832}
 \end{flushright}

\title{On Resonant Leptogenesis}

\author{Andrea De Simone $^1$ and Antonio Riotto $^{2, 3}$} 

\address{$^1$ Center for Theoretical Physics,\\
Massachusetts Institute of Technology, Cambridge, MA 02139, USA}

\address{$^2$ INFN, Sezione di Padova, Via Marzolo, 
8 - I-35131 Padua, Italy}
        
\address{$^3$ D\'epartement de Physique Th\'eorique,\\ Universit\'e de
Gen\`eve, 24 Quai Ansermet, Gen\`eve, Switzerland}
\eads{\mailto{andreads@mit.edu} and \mailto{antonio.riotto@pd.infn.it}}

\begin{abstract}
{It has been recently shown that the quantum Boltzmann equations may be 
relevant for the 
leptogenesis scenario. In particular, they lead to a 
time-dependent CP asymmetry which  depends 
upon the previous dynamics of the system. This memory effect in 
the CP asymmetry is particularly important
 in resonant leptogenesis where the asymmetry is generated by the 
decays of nearly 
mass-degenerate right-handed neutrinos.
We study the impact of the nontrivial time evolution
of the CP asymmetry in resonant
leptogenesis, both in the one-flavour case and with flavour effects 
included. We show that significant qualitative and quantitative 
differences arise with respect to the
case in which the time dependence of the CP asymmetry is neglected.  
} \end{abstract}

\maketitle

\pagestyle{fancy}
\def\thefootnote{\arabic{footnote}}
\setcounter{footnote}{0}
\setcounter{page}{1}

\fancyhead{}
\fancyfoot[C]{\thepage}


\section{Introduction}
\noindent
Thermal leptogenesis \cite{FY,leptogen,work} is 
a simple  mechanism to 
explain the observed baryon number asymmetry (per entropy density)
of the Universe  $Y_{\cal B}=\left(0.87\pm 0.02\right)
\times 10^{-10}$ \cite{wmap}.
A lepton asymmetry is dynamically generated
and then  converted into a baryon asymmetry
due to $(B+L)$-violating sphaleron interactions \cite{kuzmin,baureview}
which exist in the Standard Model (SM).
A simple model in which this mechanism can be implemented consists
of the SM plus three   right-handed (RH) Majorana neutrinos.
In thermal leptogenesis \cite{leptogen} the heavy RH neutrinos
are produced by thermal scatterings after inflation  and subsequently
decay out-of-equilibrium in a lepton number and CP-violating way, 
thus satisfying Sakharov's constraints \cite{sakharov, baureview}.
RH neutrinos  are also  a key ingredient in the formulation of
the well-known ``see-saw'' (type I) mechanism \cite{seesaw},  
which explains why neutrinos are
massive and mix among each other and  why they turn out to be
much lighter than the other known fermions
of the SM. For such a reason  thermal leptogenesis has been the
subject of intense research activity in the last few years with the main
goal of providing a quantitative
relation between the light neutrino properties and the final baryon asymmetry.

Thermal leptogenesis is based on the assumption that RH neutrinos are 
efficiently generated by thermal scatterings during the reheating stage after
inflation. In the scenario in which the RH neutrinos are hierarchical in mass,
successful leptogenesis requires that 
the RH neutrinos are heavier than about  $10^9$ GeV \cite{di}.
Hence, the required reheating temperature cannot be much lower 
\cite{leptogen,davidsonetal1,aa,jmr}. In 
supersymmetric scenarios this may be in conflict with the upper bound
on the reheating temperature necessary to 
avoid the  overproduction of gravitinos during
reheating \cite{grav}. Indeed, 
being only gravitationally coupled to the SM particles,
gravitinos may decay very late jeopardising the successful
predictions of Big Bang nucleosynthesis.

In the resonant leptogenesis scenario \cite{res, res2} this tension may be
avoided. If the RH neutrinos are nearly degenerate in mass, self-energy 
contributions to the CP asymmetries may be  resonantly enhanced, thus
making thermal leptogenesis viable at  temperatures as low as the
TeV.  Resonant
leptogenesis seems to us a natural possibility. In the absence
of simple predictive models one  expects that left-handed and RH neutrinos
show similar levels of degeneracy. Indeed, 
within the see-saw mechanism, quasi-degenerate
light neutrinos are more naturally explained by quasi-degenerate
RH neutrinos, rather than by an interplay between Yukawa couplings and the
masses of the RH neutrinos.  The quasi-degeneracy in the RH sector can be 
easily explained by  symmetry arguments, {\it e.g.} a slightly 
broken $SO(3)$ symmetry.

Resonant leptogenesis is the subject of the present paper. More specifically, 
we will investigate the resonant leptogenesis scenario
in view of the recent results achieved by studying  
the dynamics 
of thermal leptogenesis by means of 
quantum Boltzmann equations \cite{dsr} (for an earlier study, see
Ref. \cite{buchmuller}). 

Let us pause here
for a moment and summarize why quantum Boltzmann equations 
are relevant in resonant leptogenesis \footnote{For more technical 
details the reader is referred to Ref. \cite{dsr}.}.
The generation of the
baryon asymmetry occurs when RH neutrinos are out-of-equilibrium. Therefore,
their abundance and the one for the  lepton asymmetry are 
determined by Boltzmann equations. In
the classical Boltzmann equation approach,   every scattering
in the plasma is independent from the previous one and  the particle abundances
at a given time do not depend upon the previous dynamical history of the
system. 
Quantum Boltzmann equations are obtained 
starting from  the  non-equilibrium quantum field theory based on the 
Closed Time-Path  (CTP) formulation. 
It   is a powerful 
Green's function
formulation for describing non-equilibrium phenomena in field theory.  It
allows to obtain a
self-consistent set of quantum Boltzmann equations
for the quantum averages of
operators, {\it e.g.} the lepton asymmetry operator, 
 evaluated in the in-state without specifying the out-state. 
The  quantum Boltzmann equations  have an obvious interpretation in terms of 
gain and loss processes.   
What is unusual, however,   is the presence of the integral over 
 time in the scattering terms where theta functions ensure that
the dynamics is causal. The quantum Boltzmann
equations are therefore manifestly non-Markovian. Only  
the assumption that the relaxation timescale of the particle asymmetry 
is much longer than the timescale of the non-local kernels leads to 
a Markovian description. A further approximation, {\it i.e.} taking the
upper limit of the 
time integral to infinity,  leads to the familiar classical 
familiar Boltzmann  
equation. 
The physical interpretation of the integral over the past history of 
the system is straightforward: it leads to the typical ``memory'' 
effects which are observed in quantum transport theory \cite{dan,henning}. 
The thermalization 
rate obtained from the quantum transport theory may be 
substantially longer than the one obtained from 
the classical kinetic theory.

Furthermore, and more importantly, the CP asymmetry turns out to be
 a function of time, even after taking the Markovian limit. 
Its value at a given
instant depends upon the previous history of the system. 
If the time variation of the CP asymmetry is shorter than the
relaxation time of the particles abundances, the solutions to the
quantum and the classical
Boltzmann equations are expected to 
differ only by terms of the order of the ratio
of the timescale of the CP asymmetry to the relaxation timescale of the
distribution. In thermal leptogenesis with hierarchical 
RH neutrinos this is typically the case. However, in the
resonant leptogenesis scenario, at least  
two RH neutrinos $N_1$ and $N_2$ 
 are almost degenerate
in mass and the CP asymmetries from the decays of the RH neutrinos 
are resonantly enhanced if the mass difference
$\Delta M=(M_2-M_1)$
is of the order of the decay rates. The typical timescale
to build up coherently the CP asymmetry is of the order of $1/\Delta M$, which
can be larger than the timescale  
for the change of the abundance of the
RH neutrinos. This tells us that in the case of resonant leptogenesis
significant differences are expected between the classical and the
quantum approach. 

The rest of the paper is organized as follows. 
In Section \ref{1flavour} we study the impact 
of the time-dependent CP asymmetry 
on the final lepton asymmetry, in the one-flavour approximation.  
The numerical results are supported by analytical estimates,  
for both the regimes of  strong and  weak wash-out. 
The weak wash-out case is 
analyzed in greater detail since this is the regime where  
the time dependence 
of the CP asymmetry becomes more relevant. 
The generalization to more than one 
flavour is given in Section \ref{2flavours}, 
where we discuss the possible wash-out regimes.
Again, the analytical formulae confirm the numerical solutions. 
Finally, Section \ref{concl} contains our conclusions.


\section{Resonant leptogenesis revisited: the one-flavour case}
\label{1flavour}

Our starting point is the SM plus 
three RH neutrinos 
$N_{\alpha}$ ($\alpha=1,2,3$), with Majorana masses $M_\alpha$.  
The interactions among RH neutrinos, Higgs doublets $H$, lepton doublets  
$\ell_i$ and singlets $e_i$  ($i=e,\mu,\tau$) 
are described by the Lagrangian
\begin{equation}
\mathscr {L}_{\rm int}=\lambda_{\alpha i} N_\alpha
\ell_i H+h_i \bar e_i\ell_i H^c+{1\over 2}M_\alpha N_\alpha N_\alpha + {\rm h.c.}\,,
\label{lagr_flav}
\end{equation}
with summation over repeated indices.
The Lagrangian is written in the mass eigenstate basis of RH neutrinos and
charged leptons. 

For the sake of simplicity, for the rest of the paper we will restrict ourselves
to the case in which only the two lightest RH neutrinos are quasi-degenerate,
$M_1\sim M_2\ll M_3$. 
We first address the dynamics of the system in 
the so-called one-flavour approximation,  
where  Boltzmann equations
are written for the abundance of the RH neutrinos 
and  for the total lepton asymmetry. This approximation is
correct only when the interactions mediated by 
charged lepton Yukawa couplings are out of equilibrium. Supposing that
leptogenesis takes place at temperatures $T\sim M_1\sim M_2$, 
the one-flavour approximation 
holds for $M_1\gsim  10^{12}$ GeV. We will include flavour effects
later on. 

The Boltzmann equations resulting from the CTP formalism, after taking the Markovian
limit, are given by \cite{dsr} 
\begin{eqnarray}
\label{k}
Y_{N}^\prime&=&-z K\frac{\bkone(z)}{\bktwo(z)}\left(Y_{N}-Y^{\rm eq}_{N}\right)\,,
\nonumber\\
Y_{\cal L}^\prime &=& -2\,\epsilon(z)Y_{N}^\prime-\frac{1}{2}K z^3  \bkone(z)
Y_{\cal L}\,. 
\end{eqnarray}
Here $Y_{N}=Y_{N_1}\simeq Y_{N_2}$ denotes the number density of the 
RH neutrinos per entropy density which we assume to be roughly equal since
we also take the decay rates  $\Gamma_{N_1}$ and   
$\Gamma_{N_2}$ roughly 
equal;
primes stand for derivatives with respect to the ``time'' variable  $z=M_1/T$
($T$ being the temperature);
the parameter $K\equiv 
\Gamma_{N_1}/H(M_1)=\Gamma_{N_2}/H(M_1)$, where $H(z)$ is the Hubble rate, 
 controls how much RH neutrinos are
out-of-equilibrium; $Y^{\rm eq}_{N}=(1/4 g_*) z^2 K_2(z)$ is the 
equilibrium number density of RH neutrinos (being $g_*$ the number of
relativistic degrees of freedom in the plasma); 
for consistency, $Y^{\rm eq}_{N}$ has been 
computed using the Boltzmann distribution;
$\bkone$ and $\bktwo$
are the modified Bessel functions of the first and second kind, respectively. 
In Eq.~(\ref{k}) we have neglected for simplicity the contribution of
$\Delta L=1,2$ scatterings and thermal effects \cite{leptogen}.
Including
the contributions
of $\Delta L=1$ scatterings both in the wash-out term and in the CP 
asymmetry
does not change our results significantly.
We have summed up the contributions
of the two quasi-degenerate RH neutrinos employing the property
that $\epsilon_{N_1}=\epsilon_{N_2}\equiv\epsilon$. This follows from 
having assumed that the decay rates of the two RH neutrinos are 
nearly equal and therefore both CP asymmetries are resonantly enhanced. If 
the decay rates are significantly different, then one should 
pick up only the CP asymmetry contribution which is resonantly enhanced.

Finally, the time-dependent CP asymmetry relevant for
resonant leptogenesis is given by \cite{dsr}
\begin{eqnarray}
\label{kkk}
\epsilon(z)&\simeq & \overline{\epsilon}
\left[2\,
{\rm sin}^2 \left(\frac{K z^2}{4}\frac{\Delta M}{\Gamma_{N_2}}
{\Gamma_{N_2}\over\Gamma_{N_1}}
\right)
-{\Gamma_{N_2}\over \Delta M}\,
{\rm sin} \left(\frac{K z^2}{2}\frac{\Delta M}{\Gamma_{N_2}}
{\Gamma_{N_2}\over\Gamma_{N_1}}
\right)\right],\nonumber\\
\overline{\epsilon}&=& -\frac{{\rm Im}
\left[\left(\lambda\lambda^\dagger\right)^2 _{12}\right]}{\left(\lambda\lambda^\dagger
\right)_{11}\left(\lambda\lambda^\dagger
\right)_{22}}\frac{\Delta M/\Gamma_{N_2}}{1+(\Delta
M/\Gamma_{N_2})^2}.
\end{eqnarray}
The CP asymmetry therefore consists of two blocks. The first one
is the  constant piece which is the
usually adopted CP asymmetry and is resonantly enhanced for
$\Delta M=\Gamma_{N_2}$. It allows efficient generation of the lepton
asymmetry even for $M_1\sim M_2$ as low as the TeV scale. 
The other block is made of 
two oscillating functions. 
The typical timescale for the 
variation of the CP asymmetry is  

\begin{equation}
t=\frac{1}{2 H}=\frac{z^2}{2 H(M_1)}=\frac{Kz^2}{2 \Gamma_{N_1}}\sim 
\frac{1}{\Delta M}.
\end{equation} 
The CP asymmetry grows for $t\lsim 1/\Delta M$ and manifests its oscillation
pattern only for  $t\gsim 1/\Delta M$. The reader familiar with CP violation
in  neutral
meson systems would prompltly recognize that the oscillation pattern
originates
from the CP violating decays of the two mixed states $N_1$ and $N_2$.
These states do not propagate freely in the plasma though. If the 
timescale for the processes relevant
for leptogenesis is much larger  than  $\sim 1/\Delta M$, 
the CP asymmetry  should average to the constant 
value $\overline{\epsilon}$ quoted in the literature \cite{bareps, res, bareps2}. 
However, 
if the  timescale of the evolution of the CP asymmetry
is larger than or of the order of the timescale of the other processes, the 
time dependence of the CP asymmetry may not be neglected. 
This is precisely what happens in the resonant leptogenesis scenario where
$\Delta M\sim \Gamma_{N_2}\sim \Gamma_{N_1}$.

Since the strength of the
interaction 
rates is dictated by the  parameter $K$, we expect that in the strong wash-out
regime, $K\gg 1$, the effect of the time dependence of the CP asymmetry
is negligible. Due to the rapidly oscillating CP asymmetry, the lepton 
asymmetry should also rapidly oscillate and -- at large times --
reproduce the value usually quoted in the literature. 
On the contrary, in the case of weak or mild wash-out, $K\lsim 1$,
the effect of the time dependence of the CP asymmetry should be  magnified
since the CP asymmetry oscillates with a period comparable to the
time scale of the other interactions.

\begin{figure}[t]
\centering
\includegraphics[scale=0.8]{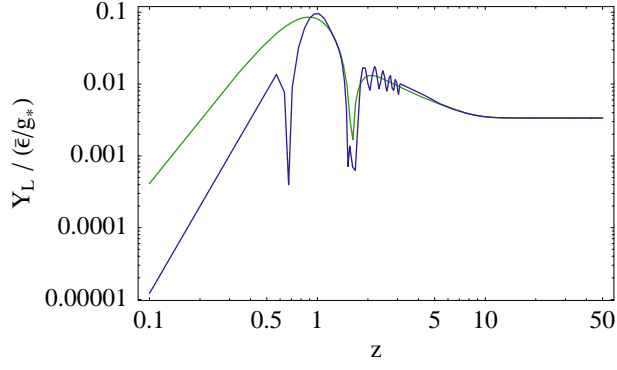}
\caption{\footnotesize The absolute value of the lepton asymmetry with the time dependence in the  CP asymmetry (blue) and without it (green), as a function of $z$, for $\Delta M/\Gamma_{N_2}=\Gamma_{N_2}/\Gamma_{N_1}=1$ and for $K=10$.}
\label{YzK10}
\end{figure}

\begin{figure}[t]
\centering
\includegraphics[scale=0.8]{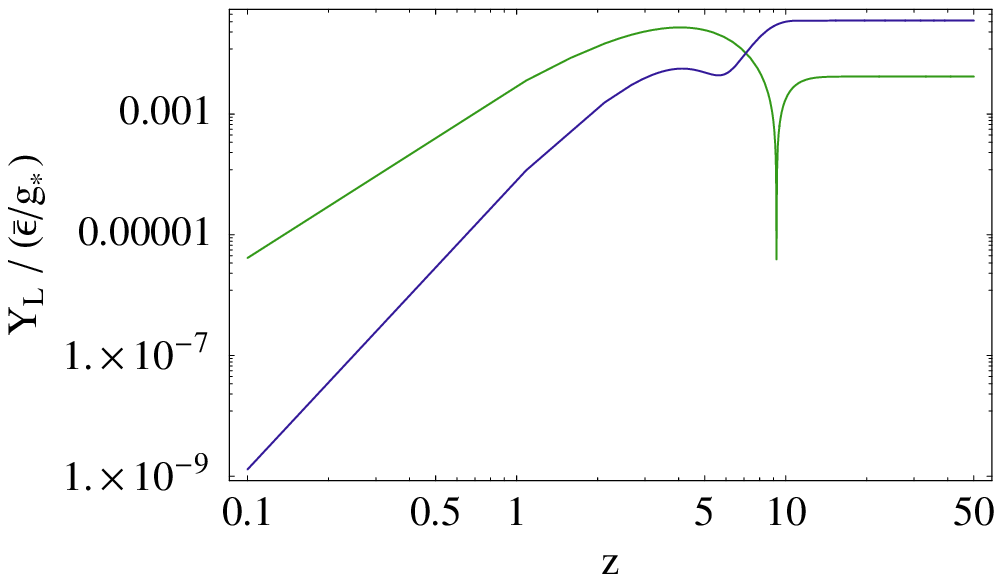}
\caption{\footnotesize  The absolute value of the lepton asymmetry with the time dependence in the  CP asymmetry (blue) and without it (green), as a function of $z$, for $\Delta M/\Gamma_{N_2}=\Gamma_{N_2}/\Gamma_{N_1}=1$ and for  $K=10^{-1}$.}
\label{YzK01}
\end{figure}

\begin{figure}[t]
\centering
\includegraphics[scale=0.8]{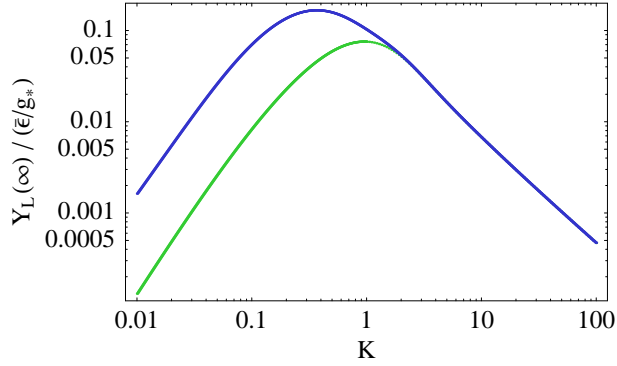}
\caption{\footnotesize The absolute value
of the final lepton asymmetry with the time dependence in the  CP asymmetry
(blue) and without (green), as a function of $K$, for $\Delta M/\Gamma_{N_2}=
\Gamma_{N_2}/\Gamma_{N_1}=1$.}
\label{YinfK}
\end{figure}

The numerical solutions of the Boltzmann equations support  these expectations. 
Figs.~\ref{YzK10} and \ref{YzK01}   show the evolution of the lepton asymmetry with and without
the time dependence in the CP asymmetry for two representative cases
of strong and weak wash-out, respectively. 
Fig.~\ref{YinfK} shows the final baryon asymmetry computed
taking into account and neglecting the time dependence in the CP asymmetry
as a function of $K$.

The numerical results can be analytically reproduced.
In the strong wash-out regime,  $K\gg 1$, $\left(Y_{N}-Y^{\rm eq}_{N}\right)
\simeq (z \bktwo(z)/4K g_*)$ and the lepton asymmetry
reads

\begin{equation}
Y_{\cal L}\simeq {1\over 2 g_*}\int_0^\infty \de z\, \epsilon(z)z^2 \bkone(z)
e^{-\frac{K}{2}\int_z^\infty \de z' (z')^3 \bkone(z')}.
\end{equation}
Using the stepeest descent method, one can easily show that the final lepton
asymmetry is equal to the one computed neglecting the time dependence
in the CP asymmetry up to small corrections 
\begin{equation}
Y_{\cal L}\simeq\frac{0.3\, \overline{\epsilon}}{g_*K^{1.16}}+
{\cal O}\left(e^{-\frac{3}{2}(\ln K)^2}
\frac{1}{K^{1/2}}\right).
\end{equation}

The weak wash-out regime is much more interesting. Let us first remind the
reader what happens in the usual case where the time dependence of the CP
asymmetry is neglected. The final lepton asymmetry results from a cancellation
between the (anti-) asymmetry generated when RH neutrinos are
initially produced and the lepton asymmetry produced when they finally 
decay. It is useful to define the value  $z_{\rm eq}\gg 1$ as the ``time'' 
when the $N$-abundance
reaches the equilibrium abundance:  
$Y_{N}(z_{\rm eq})=Y^{\rm eq}_{N}(z_{\rm eq})$. 
Since $\int_0^{z_{\rm eq}} \de z' (z')^3 \bkone(z')\simeq 
3\pi/2$, one finds that $z_{\rm eq}$ is defined implicitly by the relation
$z^{3/2}_{\rm eq}e^{-z_{\rm eq}}\simeq (3\pi K/2)$. 
For $z\lesssim z_{\rm eq}$,  inverse decays dominate over 
decays and $Y_{N}\ll Y^{\rm eq}_{N}$. From Eq. (\ref{k})
one finds 
\begin{equation}
 Y_{N}\simeq K\int_0^z \de z' z'\frac{\bkone(z')}{\bktwo(z')}Y^{\rm eq}_{N}(z')\simeq
\frac{K}{4 g_*} \int_0^z \de z' (z')^3 \bkone(z').
\end{equation}
For   $z\gsim z_{\rm eq}$, decays dominate
over inverse decays, $Y_{N}\gg Y^{\rm eq}_{N}$, and 
\begin{equation}
Y_{N}\simeq 
Y_{N}(z_{\rm eq})e^{-K\int_{z_{\rm eq}}^z \de z' z' \bkone(z')/\bktwo(z')}\simeq
Y_{N}(z_{\rm eq})e^{-K/2(z^2-z^2_{\rm eq})}. 
\end{equation}
The lepton asymmetry is therefore given by
\begin{eqnarray}
Y_{\cal L}&\simeq& -\frac{2\overline{\epsilon}K}{4 g_*}
\int_0^{z_{\rm eq}} \de z' (z')^3 \bkone(z') e^{-{K\over 2}\int_{z'}^{z_{\rm eq}} 
\de z'' (z'')^3 \bkone(z'')}\nonumber \\
&-&2\overline{\epsilon}\int_{z_{\rm eq}}^\infty \de z'\frac{\de}{\de z'}\left(
\frac{3\pi K}{8g_*}e^{-K\int_{z_{\rm eq}}^{z'} \de z'' z'' \bkone(z'')/\bktwo(z'')}
\right)\nonumber\\
&\simeq&
\overline{\epsilon}\left(\frac{1}{g_*}\left(e^{-\frac{3\pi K}{4}}-1\right)
+\frac{3\pi K}{4 g_*}\right)\simeq 
{\overline{\epsilon}\over g_*}\frac{9\pi^2}{32} K^2.
\label{without}
\end{eqnarray}

\begin{figure}[t]
\centering
\includegraphics[scale=0.8]{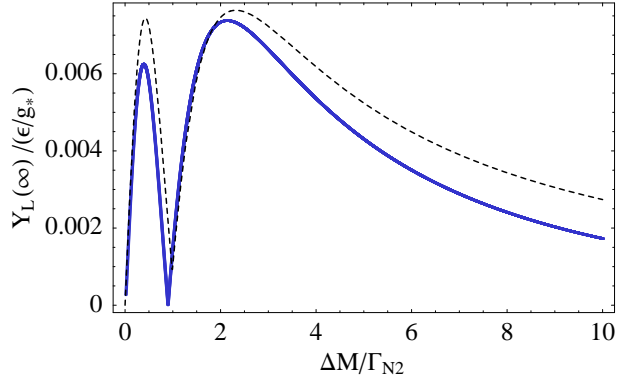}
\caption{\footnotesize  The absolute value
of the final lepton asymmetry as a function of $\Delta M/\Gamma_{N_2}$ 
for $K=10^{-2}$, $\Gamma_{N_2}/\Gamma_{N_1}=1$.
The blue solid line is obtained by numerically integrating the 
Boltzmann equations; the black dashed line represents the analytical approximation
given in the text. The normalization of the asymmetry by $(\epsilon/g_*)$ 
has been performed only with the part of $\epsilon$ which is independent of $\Delta M/\Gamma_{N_2}$, so that the full dependence of $Y_{\cal L}$ on $\Delta M/\Gamma_{N_2}$ is explicitly shown.}
\label{YnewK001}
\end{figure}

Let us now come back to the case in which the time dependence
of the CP asymmetry is accounted for. This time the near-cancellation
between the (anti-) asymmetry generated when RH neutrinos are
initially produced and the lepton asymmetry produced when they finally 
decay is not expected to hold since the asymmetry is modulated by
the varying CP asymmetry. Again, we split the final lepton asymmetry as the
sum of two contributions

\begin{eqnarray}
Y_{\cal L}&\simeq& \frac{2\overline{\epsilon}K^2}{8 g_*}
\int_0^{z_{\rm eq}} \de z' (z')^5 \bkone(z') 
+\frac{3\pi K^2}{4 g_*}\int_{z_{\rm eq}}^\infty \de z' 
z' \epsilon(z')
e^{-K/2(z^2-z^2_{\rm eq})}
\nonumber\\
&\simeq&
\frac{\overline{\epsilon}}{g_*}
\left[\frac{45\pi K^2}{8}+\frac{3\pi K}{4} \frac{1}{
1+\left(\Delta M/\Gamma_{N_2}\right)^2}\left(1+
\left(\frac{\Delta M}{\Gamma_{N_2}}\right)^2\right.\right.\nonumber\\
&&-2\left.\left.
\cos\left[\frac{K z_{\rm eq}^2}{2} \frac{\Delta M}{\Gamma_{N_2}}
\right]+\left(\frac{\Delta 
M}{\Gamma_{N_2}}-\frac{\Gamma_{N_2}}{\Delta 
M}\right)\sin\left[\frac{K z_{\rm eq}^2}{2} \frac{\Delta 
M}{\Gamma_{N_2}}
\right]\right)\right],\nonumber\\
&&
\label{with}
\end{eqnarray}
where in the first contribution we have approximated
the CP asymmetry
by $\epsilon(z)\simeq -\overline{\epsilon}(Kz^2/2)$. 
This formula reproduces fairly well the complicated pattern 
shown in Fig.~\ref{YnewK001}
and shows that the resonance is displaced from the position 
 $\Delta M/\Gamma_{N_2}=1$ obtained when the time dependence
of the CP asymmetry is neglected. 
Similarly to Eq. (\ref{without}), the lepton asymmetry scales like $K^2$. 
However, this scaling is not due to cancellations among the
asymmetries at different stages, but rather to the 
fact that for tiny values of $K$, or equivalently for $t\lsim 1/\Delta M$, 
the CP asymmetry is suppressed. 

\begin{figure}[t]
\centering
\includegraphics[scale=0.8]{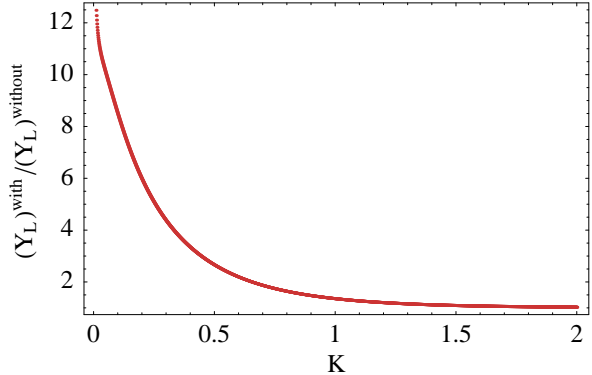}
\caption{\footnotesize The ratio of the absolute value
of the final lepton asymmetry with and without 
the time dependence in the  CP asymmetry as a function of $K$,  
for $\Delta M/\Gamma_{N_2}=\Gamma_{N_2}/\Gamma_{N_1}=1$.}
\label{DM1R1}
\end{figure}

For $\Delta M/\Gamma_{N_2}=1$, Eq. (\ref{with}) reduces to
\begin{equation}
Y_{\cal L}\simeq 
\frac{\overline{\epsilon}}{g_*}
K\left[\frac{45\pi}{8}K+\frac{3\pi}{2}\sin^2
\left(\frac{K z_{\rm eq}^2}{4}\right)
\right].
\label{with1}
\end{equation}
For example, for 
$K=10^{-2}$, 
the ratio between the final lepton asymmetries
given in Eqs.~(\ref{with1}) and (\ref{without}) is  $\sim  11$, 
which is in good agreement with our numerical results, 
as shown in Fig.~\ref{DM1R1}.

\section{Resonant leptogenesis revisited: the flavoured case}
\label{2flavours}

Let us now go beyond the one-flavour approximation.
As we have already mentioned, it is
rigorously correct only when the interactions mediated by 
charged lepton Yukawa couplings are out of equilibrium. Supposing that
leptogenesis takes place at temperatures $T\sim M_1\sim M_2$, 
the one-flavour approximation 
only holds for $M_1\sim M_2 \gsim 10^{12}$ GeV. In this range all
the interactions mediated by  the 
charged lepton Yukawa couplings are out of equilibrium and there is no notion
of flavour. One is allowed to perform a rotation
in flavour space to store all the lepton asymmetry in one flavour, the
total lepton number. 
However, at $T\sim 10^{12}$ GeV, the interactions mediated by the
charged tau Yukawa come into equilibrium followed by those mediated
by the charged muon Yukawa at $T\sim 10^{9}$ GeV and the
notion of flavour becomes physical.
Including the issue of flavour can  significantly affect the result for the
final baryon asymmetry  
\cite{barbieri,flavour, pil, nardi,davidsonetal1,davidsonetal2,flavourextra}. 
Thermal leptogenesis is a dynamical
process, involving  the  production and 
destruction of RH neutrinos 
and  of the  lepton asymmetry  that is distributed among
distinguishable flavours.  The processes which
wash out lepton number
are flavour dependent, {\it e.g.} the inverse decays
from electrons can destroy the lepton asymmetry carried by,
and only by,  the electrons.
When flavour is accounted for, 
the final value of the baryon asymmetry is the sum of three
contributions. Each term is given by the 
CP asymmetry in a given flavour    properly weighted
 by a wash-out factor induced by the    
lepton number  violating processes for that flavour.  

The Boltzmann equations
with flavour taken into account are\footnote{We neglect  both quantum 
flavour
correlations \cite{davidsonetal1} and the corrections
arising from connecting the asymmetries in the lepton doublets to the
asymmetries in the charges $\Delta_i=B/3-L_i$ conserved by weak sphalerons
\cite{barbieri,davidsonetal1,nardi,davidsonetal2}. They introduce small
corrections to our results.}
\begin{eqnarray}
\label{kflavour}
Y_{N}^\prime&=&-z K\frac{\bkone(z)}{\bktwo(z)}\left(Y_{N}-Y^{\rm eq}_{N}\right),
\nonumber\\
Y_{{\ell}_i}^\prime&=& -2\,\epsilon_i(z)Y_{N}^\prime-\frac{1}{2}K_i z^3  \bkone(z)
Y_{\ell_i}\,, \qquad (i=e,\mu,\tau), 
\end{eqnarray}
where  
\begin{eqnarray}
\epsilon_i(z)&\simeq & \overline{\epsilon}_i
\left[2\,
{\rm sin}^2 \left(\frac{K z^2}{4}\frac{\Delta M}{\Gamma_{N_2}}
{\Gamma_{N_2}\over\Gamma_{N_1}}
\right)
-\frac{\Gamma_{N_2}}{\Delta M}\,
{\rm sin} \left(\frac{K z^2}{2}\frac{\Delta M}{\Gamma_{N_2}}
{\Gamma_{N_2}\over\Gamma_{N_1}}
\right)\right],\nonumber\\
\overline{\epsilon}_i&=& -\frac{
\sum_{j=e,\mu,\tau}{\rm Im}
\left(\lambda_{1i}
\lambda_{1j}\lambda^\dagger_{j2}
\lambda^\dagger_{i2}\right)
}{\left(\lambda\lambda^\dagger
\right)_{11}\left(\lambda\lambda^\dagger
\right)_{22}}\frac{\Delta
M/\Gamma_{N_2}}{1+(\Delta
M/\Gamma_{N_2})^2},\nonumber\\
K_i&\equiv&\frac{ 
\Gamma\left(N_1\rightarrow \ell_i H\right)}{H(M_1)}\simeq
\frac{
\Gamma\left(N_2\rightarrow \ell_i H\right)}{H(M_1)}, 
\,\,\, K=\sum_{j=e,\mu,\tau}K_j. 
\end{eqnarray} 
Notice  that, for simplicity,  we have again assumed that
the partial rates for the decay of the two quasi-degenerate RH neutrinos
are nearly equal. Extending our study to the case
where the rates are different is straightforward. 

\begin{figure}[t]
\centering
\includegraphics[scale=0.8]{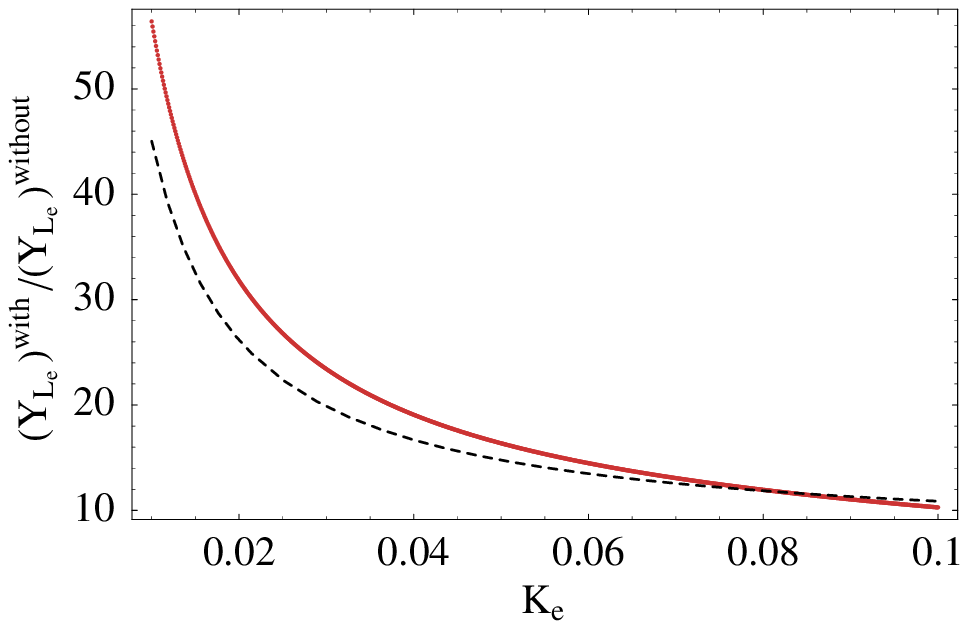}
\caption{\footnotesize The ratio of the absolute value
of the final lepton asymmetry in the $e$ flavour, with and without 
the time dependence in the  CP asymmetry, as a function of $K_e$, 
for $K_\mu=0.05$, $\Delta M/\Gamma_{N_2}=\Gamma_{N_2}/\Gamma_{N_1}=1$.
The red solid line is obtained by numerically integrating the 
Boltzmann equations; the black dashed line represents the analytical approximation
given in the text.}
\label{Ki-small}
\end{figure}

What we have learned so far is that 
significant differences with respect to the case in which the CP asymmetries
are constant in time are expected in the weak wash-out regime. Let us then 
suppose
that all flavours are in the weak wash-out regime and also that
$K\lsim 1$. If the time dependence of the CP asymmetry is neglected each
individual flavour asymmetry suffers the usual cancellation
and is  given by $Y_{\ell_i}\simeq 2.8
(\overline{\epsilon_i}/g_*)KK_i$ \cite{davidsonetal2}. 
Solving the Boltzmann equations along the
same lines  leading
to  the results (\ref{with}) and (\ref{with1}) shows that 
flavour asymmetries, when the time dependence in $\epsilon_i$ 
is taken into account, are given by (for $\Delta M/\Gamma_{N_2}=1$)
\begin{equation}
Y_{\ell_i} 
\simeq
\frac{\overline{\epsilon}_i}{g_*}
K\left[\frac{45\pi}{8}K+\frac{3\pi}{2}\sin^2
\left(\frac{K z_{\rm eq}^2}{4}\right)
\right].
\label{qq}
\end{equation}
Therefore, the flavour asymmetries
are increased by a factor $\sim 10(K/K_i)$ compared to what is obtained
when the CP asymmetry is assumed to be a constant. 
This result is confirmed by the numerical study shown in Fig.~\ref{Ki-small}.

\begin{figure}[t]
\centering
\includegraphics[scale=0.8]{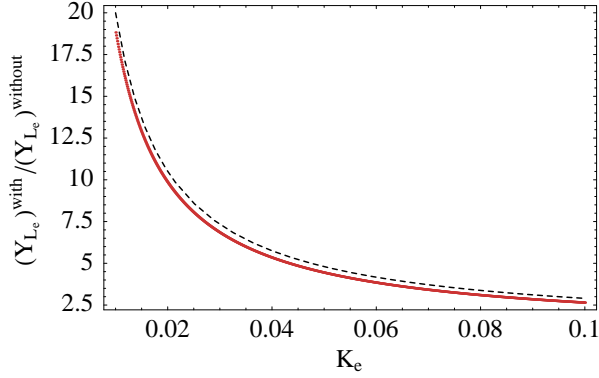}
\caption{\footnotesize The ratio of the absolute value
of the final lepton asymmetry in the $e$ flavour, with and without 
the time dependence in the  CP asymmetry, as a function of $K_e$,
for $K_\mu=10$, $\Delta M/\Gamma_{N_2}=\Gamma_{N_2}/\Gamma_{N_1}=1$.
The red solid line is obtained by numerically integrating  the 
Boltzmann equations; the black dashed line represents the analytical approximation
given in the text.}
\label{Ki-mixed}
\end{figure}

Let us now consider the case in which some flavour is in the
strong wash-out regime, so that $K\gg 1$, but some flavour $i$ is weakly 
coupled, $K_i\lsim 1$. Notice that this case is peculiar to the scenario where
flavour is taken into account; it has no analogue in the one-flavour regime.
Now the lepton asymmetry stored in the 
flavour $i$ without accounting for the time-dependence of the CP asymmetry
reads 
\cite{davidsonetal2}
\begin{eqnarray}
Y_{\ell_i}&\simeq& -2\overline{\epsilon}_i
\int_0^\infty \de z Y_N^\prime
e^{-\frac{K_i}{2}\int_z^\infty \de z' (z')^3 K_1(z')}\nonumber\\
&=&\overline{\epsilon}_i K_i\int_0^\infty \de z
Y_N^{\rm eq} z^3 \bkone(z)\simeq 0.8\frac{\overline{\epsilon}_i
  K_i}{g_*}
\label{mixedold}
\end{eqnarray}
where we have made use of the fact that, for $K\gg 1$, $Y_N\simeq Y_N^{\rm eq}$.
The result is proportional to the weak wash-out parameter $K_i$
because, in the limit of vanishing $K_i$, the asymmetry produced
by inverse decays is cancelled by the one generated by the decays
of the RH neutrinos. 
If the time dependence of the CP asymmetry is accounted for, the lepton
asymmetry
in the flavour $i$ is given by
\begin{eqnarray}
Y_{{\ell}_i}&\simeq& -2
\int_0^\infty \de z \epsilon_i(z) Y_N'
e^{-\frac{K_i}{2}\int_z^\infty \de z' (z')^3 \bkone(z')}\nonumber\\
&\simeq &  2\int_0^\infty \de z \epsilon_i^\prime(z) Y_N
+K_i\int_0^\infty \de z  \epsilon_i(z) Y_N z^3 \bkone(z).
\label{mixednew1}
\end{eqnarray}
where in the second line we have neglected the damping 
factor. The piece proportional to the derivative of the CP asymmetry may
be evaluated as follows. Since $K\gg 1$, the RH abundance can be
approximated as
\begin{equation}
Y_N(z)\simeq \left\{\begin{array}{lc}
\frac{1}{2 g_*}\left(1-e^{-\frac{K}{6}z^3}\right) & (z<z_{\rm eq})\,,\\
\frac{1}{4 g_*}z^2 \bktwo(z)& (z>z_{\rm eq})\,,
\end{array}\right.
\label{YNK}
\end{equation}
where $z_{\rm eq}\simeq (6/K)^{1/3}$. The integral can now be evaluated  
as the sum of three different pieces, for $0<z<z_{\rm eq}$, $z_{\rm 
eq}<z<1$ and for $z>1$. The latter is negligible since for large $K$ the
oscillating functions in the derivative of the CP asymmetry average to 
zero. The other two integrals can be easily evaluated using the expression
(\ref{YNK}) and give
\begin{equation}
\fl
\hspace{1cm}
2\int_0^\infty \de z \epsilon_i^\prime(z) Y_N\simeq
-\frac{\overline{\epsilon}_i}{2 g_*}z_{\rm eq}
\frac{\Gamma_{N_2}}{\Delta M}\left[\frac{\Gamma_{N_2}}{\Delta M}\cos\left(
\frac{K z^2_{\rm eq}}{2}\frac{\Delta M}{\Gamma_{N_2}}\right)-
\sin\left(
\frac{K z^2_{\rm eq}}{2}\frac{\Delta M}{\Gamma_{N_2}}\right)
\right].
\end{equation}
The final lepton asymmetry in the flavour $i$ is therefore
\begin{equation}
\fl\hspace{1.5cm}
Y_{{\ell}_i}\simeq 0.8\frac{\overline{\epsilon}_i}{g_*}K_i-
\frac{\overline{\epsilon}_i}{2 g_*}z_{\rm eq}
\frac{\Gamma_{N_2}}{\Delta M}\left[\frac{\Gamma_{N_2}}{\Delta M}\cos\left(
\frac{K z^2_{\rm eq}}{2}\frac{\Delta M}{\Gamma_{N_2}}\right)-
\sin\left(
\frac{K z^2_{\rm eq}}{2}\frac{\Delta M}{\Gamma_{N_2}}\right)
\right].
\label{mixednew}
\end{equation}
For instance, for $K=10$ and $\Delta M\sim \Gamma_{N_2}$, one obtains 
$z_{\rm eq}\simeq 0.85$ and the 
 ratio between the asymmetries (\ref{mixednew})  
and  (\ref{mixedold})  goes like $1+(0.19/K_i)$, which agrees with the
numerical results,  as shown in Fig.~\ref{Ki-mixed}.


\section{Conclusions}
\label{concl}
\noindent
Resonant leptogenesis has received much attention 
since it allows an efficient generation of the baryon asymmetry for RH 
neutrinos as light as the TeV scale. Through 
a non-equilibrium quantum field theory
approach to leptogenesis, we have recently  shown that 
the CP asymmetry parameter is not constant in time, but it varies
with a typical timescale equal to the mass difference 
of the RH neutrinos \cite{dsr}. 
In 
resonant leptogenesis, the  
two RH neutrinos $N_1$ and $N_2$ 
 are almost degenerate
in mass and the CP asymmetry from the decay of the first RH neutrino $N_1$ 
is resonantly enhanced if the mass difference
$\Delta M=(M_2-M_1)$
is of the order of the decay rate of the RH neutrinos. 
Therefore, the  typical timescale of variation of the
CP asymmetry 
can be larger than the timescale  
for the change of the abundances of the
RH neutrinos. 

We have studied what differences arise with respect to the case in which
the CP asymmetry is assumed to be a constant. Let us summarize our results:

1)  One-flavour case, valid for
$M_1\sim M_2\gsim 10^{12}$ GeV:  the expression for the 
final baryon asymmetry differs
from the results appeared so far in the litarature in the weak 
and mild wash-out regime, $K\lesssim 1$. The baryon asymmetry scales
like  $K^2$, see Eqs.~(\ref{with}) and (\ref{with1}), and is a factor
${\cal O}(10)$ larger than the baryon asymmetry computed with constant
CP asymmetry.

2) Flavoured case,  to be applied when 
$M_1\sim M_2\lesssim 10^{12}$ GeV:  the expression for the 
final baryon asymmetry differs
from the results appeared so far in the litarature for those
flavours $i$ which are in the  weak 
and mild wash-out regime. If $K\lesssim 1$ and $K_i\lesssim 1$, the 
asymmetry in the flavour $i$ scales like $K^2$, see Eq.~(\ref{qq}),  and is 
enhanced by a factor $\sim 10(K/K_i)$ with respect to the case in which
the CP asymmetry is constant. If  $K\gsim 1$ and $K_i\lesssim 1$, the 
asymmetry in the flavour $i$ is given by  Eq.~(\ref{mixednew}) and can be
enhanced by a 
 factor proportional to $1/K_i$  with respect to the case in which
the CP asymmetry is constant.

We conclude that the memory effects encoded in the time-dependent CP asymmetry
may play an important role in resonant leptogenesis.

\ack

A.D.S. is supported in part by  INFN 
``Bruno Rossi'' Fellowship and in part by 
the U.S. Department of Energy (D.O.E.) 
under cooperative research agreement DE-FG02-05ER41360.


\end{document}